# Leveraging Generative AI for large-scale prediction-based networking

Mathias D. Thorsager, Israel Leyva-Mayorga, *Member, IEEE*, and Petar Popovski, *Fellow, IEEE*

*Abstract*—The traditional role of the network layer is to create an end-to-end route, through which the intermediate nodes replicate and forward the packets towards the destination. This role can be radically redefined by exploiting the power of Generative AI (GenAI) to pivot towards a prediction-based network layer, which addresses the problems of throughput limits and uncontrollable latency. In the context of real-time delivery of image content, the use of GenAI-aided network nodes has been shown to improve the flow arriving at the destination by more than 100%. However, to successfully exploit GenAI nodes and achieve such transition, we must provide solutions for the problems which arise as we scale the networks to include large amounts of users and multiple data modalities other than images. We present three directions that play a significant role in enabling the use of GenAI as a network layer tool at a large scale. In terms of design, we emphasize the need for initialization protocols to select the prompt size efficiently. Next, we consider the use case of GenAI as a tool to ensure timely delivery of data, as well as an alternative to traditional TCP congestion control algorithms.

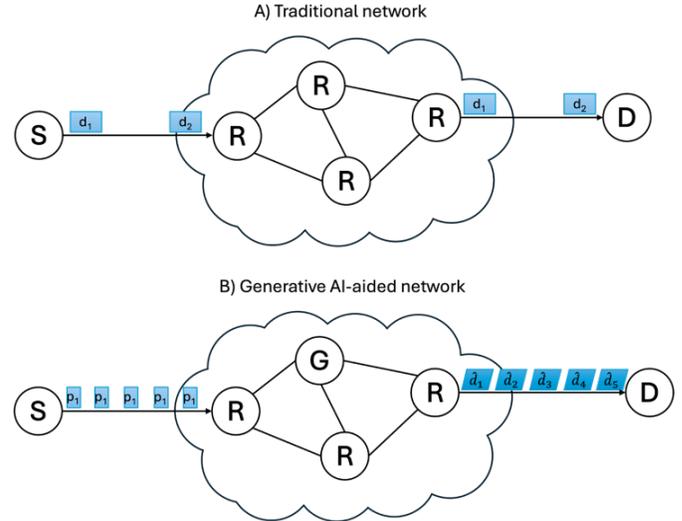

**Fig. 1.** A) Under the traditional view of a communication network, the nodes act as simple relays that replicate the data hop-by-hop. B) Adding generative AI to the network enables it to exploit prediction to increase the amount of data delivered end-to-end (E2E) beyond classical measures of capacity.

## I. INTRODUCTION

Shannon defined information as a measure of uncertainty. Then, it follows that the purpose of communication is to transmit the data residing at the source that corresponds to the amount of uncertainty the destination has about the data at the source. Henceforth, the part of the data that can be predicted is not informative and should not be communicated. Despite this fundamental principle, digital communication systems operate under the premise of replication and the assumption that all data is informative. Consequently, the aim of networking protocols is to ensure a replica of the data created at a source node arrives at the destination node. Thus, networks typically behave as dumb pipes, with relay nodes that simply replicate the data, hop by hop, until it reaches the destination. The role of network nodes was generalized by network coding [1], where network nodes act as encoders combining packets, possibly combining multiple traffic flows. By following this behavior, a maximum network capacity can be found, called the maximum flow, which is measured in bits per channel use, that can never be exceeded, simply because the links that are used to deliver the data end-to-end (E2E) have a finite capacity.

Nevertheless, prediction at the source can be used to improve the communication process. For instance, caching [2] is used in multiple applications nowadays, such as video streaming, and relies on prediction of the data that is likely to be relevant for the destination. This data is transmitted from the source to an edge node preemptively, which stores it until requested by the destination. Here, while both the edge node and the destination receive replicas of the packets generated by the source, the data does not need to be delivered E2E in real-time. Instead, the data is only delivered from the edge node to the destination, which provides an instantaneous illusion of increased E2E network capacity and reduced latency. Besides, it is only efficient in terms of communication and storage resources when the system can accurately predict the data that will be requested at a later time.

The next leap in network prediction is exploiting Generative AI (GenAI) for real-time networking. In our previous work [3], we explored the use of GenAI in an innovative way that allows for sidestepping the capacity limitations in a communication network. Such limitations in traditional packet replication approaches occur due to the need to preserve the amount of incoming and outgoing data at every node that acts as a relay between the source and the destination nodes. Nevertheless, we illustrate how GenAI can be used to increase the capacity of a network by leveraging GenAI nodes close to the destination node. This novel networking paradigm allows to dramatically reduce the amount of data transmitted from the source to the GenAI node, which then augments the data that arrives at the destination. This is illustrated in Fig. 1 where the constant flow in the traditional network is changed by transmitting smaller versions of the data (prompts) and generating approximations of the original data which is received at the destination. Thus, the network capacity is increased by sidestepping the limitations in the source-to-GenAI path and transmitting at maximum capacity from the GenAI node onwards. These gains were exemplified in [3] through an analysis of image generation where the use of a popular image generation model [4] improved the flow through the network by more than 100% when compared to a baseline JPEG compression scheme.

While image generation is a common application for GenAI, we envision the possibilities of exploiting GenAI for many other data modalities. As such, to realize the gains shown in [3] as we start generalizing the scheme to new data types as well as scaling the network to include large numbers of devices and nodes, we need to develop several protocols which tackle the new problems that arise. Resource allocation will be of particular importance, since an increase in the number of devices will place a great computational load on the GenAI-aided network nodes. Thus, to reduce the end-to-end latency, resource allocation mechanisms must consider the load on the nodes, their computational capability, and the amount of time required for their models to generate the data and pair the sources with appropriate network nodes.

## II. SURVEY

In the context of GenAI, foundation models have been developed to operate based on different inputs known as *prompts* which, for the purposes of the generative network layer scheme, can be categorized into explicit and implicit prompts, as illustrated in Fig. 2. For the following brief overview of these prompting types, we mainly focus on the foundation models for media (image and video) generation and Natural Language Processing (NLP). However, we note that the principles of models for image generation are being applied for other data modalities such as audio [5]. For the media generation based on explicit prompts, two main technologies are used in the current literature: Generative Adversarial Networks (GAN) and Denoising Diffusion Probabilistic Models (DDPM). Most of these models are based on the text-to-image or text-to-video prompt style where the content is generated based on explicit text prompts which provide an abstract textual description of the desired content [6][7][8][9]. While these models can create content with an exceedingly high perceptual quality, these models may face limitations when tasked with creating close approximations of existing content due to the abstract descriptions in the prompts. This is mitigated by changing the prompts to consist of latent representations or embeddings of the images, which retain more information about the original images than a textual prompt at the cost of a larger prompt [4][10][11]. With implicit prompts, media generation mainly consists of the prediction of future frames based on the history of previous frames or the interpolation of intermediate frames [12][13][14]. As the generation of frames is based on previous (and future) frames of the same content, likely showing the same subjects and scenery, models using implicit prompts are more likely to generate content with a low distortion. However, when the content consists of still images where no prior or otherwise similar images exist implicit prompting becomes problematic to use. LLM models that generate text based on text prompts, also known as text-to-text models, fall somewhere between implicit or explicit prompting. Regardless of the desired prompting style, the prompt itself consists of a textual prompt. If the text prompt instructs the model to generate a paragraph based on some abstract description of the desired content, the prompt is seen as an explicit prompt. However, if the prompt is a series of data points (i.e. sensor measurements), and the model has previously been instructed to predict the next step of the series, the prompt is an implicit prompt. As such, the distinction of the prompt style lies not with the prompt modality, but purely with the intent of the prompt and with how the model has been instructed to interpret the prompt. This also means that any recent LLM such as OpenAI's GPT-4 [15], Google's BARD [16], and Apple's MM1 [17] can handle prompts of both types. However, when considering the use of LLMs in the context of the generative network layer scheme, the flow gain which can be expected when using explicit prompting is very limited. Besides the issue, which also pertains to the media generation models, of accurately creating content that is similar to the original, the difference in the size of the prompt and the generated content is likely relatively small. It is therefore likely only feasible to use implicit prompting with LLMs. Here we put emphasis on the LLM's ability to understand multimodal data which makes a versatile tool for use in telecommunication [18]. Furthermore, LLMs may see better use as content regulators where the LLMs are used to understand the content and network conditions to control network parameters more accurately when delivering the content [19][18][20][21].

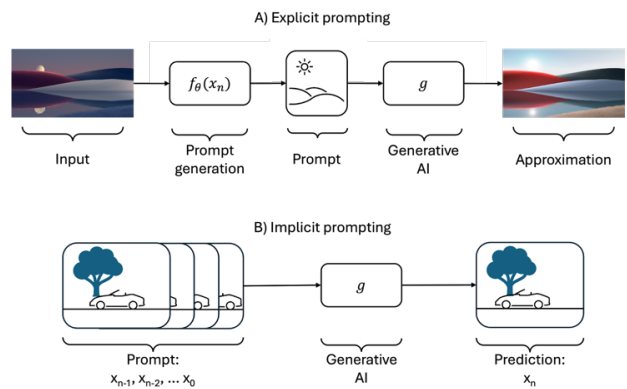

**Fig. 2.** Generative AI can operate under explicit prompting and implicit prompting. A) In the former, the prompt is created by extracting the relevant features of the original data and the goal of generative AI is to approximate the original data based on the prompt. B) In the latter, the prompt is constructed with original data, which serves as context for the generative AI to perform predictions.

For some years, the focus in image generation models was on creating the most realistic images while accurately depicting what was described in the prompt. As text-to-image models matured and improved in this aspect, some of the focus started shifting to improving other aspects of the generation process, mainly focusing on the generation time [7][22][23]. The improvements made to generation time have indeed made these models more approachable for use in networking, reaching as low as 9 ms for a single image generation. However, currently these gains do come at a slight cost of image quality. In the same line, generative compression models have mostly focused on lowering the size of the prompts as much as possible while maintaining a high perceptual quality. We also now see the same trend of investigating generative models for generative compression which decrease both the encoding and decoding time though, again, at a slight cost of image quality. In the current state of the art for generative compression models, the generation time is in the tens of milliseconds [24][25].

## III. VISION

Recently, GenAI has made its way into numerous applications such as text and image generation, but the end consumers of the data have so far been humans. Instead, as illustrated in Fig. 3, we envision that the end consumers of the generated data in 6G and following systems will also include network nodes and other devices, such as robots and IoT nodes. In other words, we are still in the machine-to-human stage of GenAI but we will soon move to the machine-to-machine stage, which requires mechanisms to support this transition and get its benefits. Specifically, the suitability of GenAI for real-time and latency-constrained networking is still largely unexplored. This is partly because GenAI is a new concept and the generation of multimedia content still takes a considerable amount of time, even in today's most powerful computing nodes. While a generation time of 50 ms per image [24] may not be significant on the scale of individual users, as the number of users increases the resources of the computing nodes must be split among them. Therefore, introducing GenAI in real-time networking brings up new challenges on the timing aspects, but also on the metrics to measure the quality of the generated data for the application at hand. For instance, traditional distortion metrics such as the Mean Squared Error (MSE) can be used, but these only capture the similarity between the original data and the generated data and do not account for the loss in performance at the consumer of the data, which can be a user or an application. Moreover, future networks will employ GenAI on user content, on control information, protocol signaling, or sensing data, and the features that are relevant for each of these cases are widely different. This caters for a semantic and goal-oriented view of the communication and generation process as a whole, where the amount of transmitted and generated data is optimized to achieve a high-level objective set by the end consumer of the data. In the following, we present three interesting research directions to maximize the benefits of GenAI for networking.

### A. Initialization Protocol for Generative Network Layer

An essential step in optimizing the generation and transmission of different types of data and data used for different purposes is determining the relevant function which describes the quality of the data. Even in the case of the image transmission investigated in [3], we cannot always assume that the transmitter knows the appropriate rate-quality function for all images. We have observed that images with certain contents behave differently resulting in unique quality distributions for the same prompt sizes. As such, to optimally use GenAI to aid in the transmission of data, an initialization process should be carried out where the transmitter learns the specific quality function for the particular data it is going to transmit. Besides being the key to enabling the use of GenAI in the network layer, a data-specific initialization protocol can help increase the efficiency of the generative network layer scheme. Depending on the granularity with which data is differentiated and on the quality function that is used, the transmitter may be more or less confident of the exact quality of the generated content received at the destination. Take again the example case of image transmission and assume that all images can be categorized into two classes: 1) images with people and 2) images without people. Also consider the intuitive assumption that images with people are more difficult to generate, meaning that the quality is lower for a given prompt size. If, for each individual image to be transmitted, the transmitter is not aware of the class it belongs to, it will have to consider that it may be an image containing a person and choose, preemptively, a larger prompt size to meet the required quality at the receiver. This does, however, mean that the transmitter 1) must be able to discern the classes from which the data to transmit can be selected and 2) must change its transmission strategy accordingly.

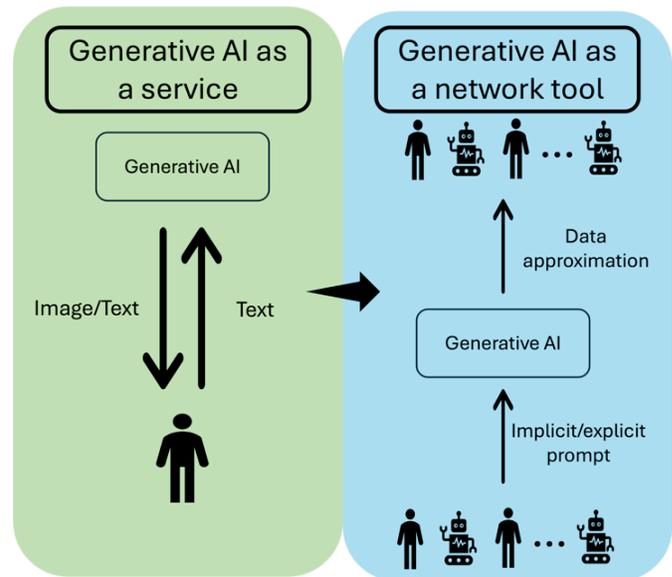

**Fig. 3.** Envisioned evolution from generative AI as a service for human-to-human communication and entertainment towards a powerful networking tool.

## B. Implicit Prompting for Timely Delivery of Data

Depending on the type of data transmitted through the GenAI aided network nodes, using implicit prompting might be a better alternative to explicit prompting. That is, by considering the previously received data as the prompt, as shown in Fig. 2, the GenAI-aided network node can predict future data. Then, it follows that by artificially dropping data packets at the transmitter and relying on the generation of these packets, it is possible to increase the effective flow over constrained links similar to what we have shown for the explicit prompting. Nevertheless, the potential of implicit prompting for ensuring the timely delivery of periodic data seems to be even greater. In cases of mobile networks where the channel conditions are constantly changing, we may experience sporadic (and possibly correlated) channel degradation, resulting in dropped packets. In these cases, a GenAI model can monitor the flow of the periodic data and, in cases of delayed packets, the GenAI model can send a generated approximation of the data to ensure that the receiver receives all the data on time. The main challenge of using GenAI to ensure the timely delivery of data is the delay introduced by the generation of the data. While this will inevitably change depending on the type of data being predicted, it seems counterintuitive to solve a timing problem with a tool which introduces a significant delay. However, predictive GenAI models can predict more than a single time step ahead, which allows them in this case to start generating data several periods before it is expected to be received. With this, when a packet is delayed, the GenAI-aided network node already has a prediction ready to transmit, removing any additional delay. On the downside, the further ahead a GenAI model predicts, the worse the quality of the prediction will be. For most data types this is quite intuitive, the further ahead a model must predict the larger the uncertainty of the true state becomes. There is an additional dimension to consider for predictive multimedia GenAI models, as the perceptual quality of the media degrades with the number of consecutively generated steps, in addition to the degradation due to trying to predict the movement of an object further into the future. This means that there is an inherent tradeoff between both the prediction accuracy and induced delay as well as the perceptual quality and induced delay.

## C. Generative Network Layer for TCP Congestion Control

Parallels can be drawn between the generative network layer scheme and TCP Congestion Control Algorithms (CCA) in the transport layer. Both aim to overcome deficiencies in link capacity by altering the transmission strategy. As such, we envision the possibility of utilizing GenAI as a tool in TCP CCA. The simple strategy is to directly replace the use of Contention Windows (CWnD) in TCP CCA algorithms and instead control the prompt size of the data being transmitted. While both approaches achieve the same goal of lowering the flow over a link, by controlling the prompt size, we maintain a consistent rate of data received at the destination(s). We can extend this approach by using a combination of the CWnD and controlling the prompt size to maintain greater control over several cases of link capacities. The main case where we envision that the use of GenAI can aid the TCP CCA is when high-capacity links experience sporadic congestion. Commonly used TCP CCAs are designed to quickly lower the CWnD in case of detected congestion and slowly increase the CWnD to probe the capacity of the network. This means that, in the case of short bursts of congestion, the device will spend a prolonged amount of time transmitting at a lower rate than what is possible [26]. In these cases, the TCP CCA could choose to lower the size of prompts instead of the CWnD, but this comes with the challenge of determining the duration of congestion. As such, we can think of an alternative which can achieve the same result but from the perspective of the relay nodes in the network.

Network congestion due a sudden burst of traffic results in a sudden increase in the queue size of a relay node. The most commonly used TCP CCAs (Reno, CUBIC, etc.) only use acknowledgements as a metric to determine the CWnD. However, using only this metric means that the devices are unable to detect increasing queue sizes at the relay nodes. They will only detect the congestion when a timeout is triggered by not receiving an ACK for a certain amount of time. This means that they will continue sending at the same rate for some amount of time, exacerbating the problem of queue length. Furthermore, in traditional TCP CCA, only the source can control the amount of data being sent to and through the network. However, with GenAI we can enable the network relays to dynamically adjust the size of the data by generating prompts on behalf of the source. When the relay nodes detect the sudden increase in queue size, they can choose to generate prompts of the data in the queue to artificially lower the size of the queue. If the congestion is short lived, the load on the problematic node will eventually decrease and the node can go back to sending the original data. If, however, the congestion period is too long, then they can still go back to sending the original data to allow the devices to detect the prolonged congestion and lower their CWnD. As such, using GenAI for this purpose will either allow the devices to continue transmitting without noticing the congestion or simply delay the detection of congestion without increasing the severity of the congestion.

## IV. FINAL THOUGHTS

Generative AI (GenAI) is a promising tool for the network layer which has been shown to enable a significant increase in network flow when transmitting images. However, networks with large amounts of users will require resource allocation schemes to solve the problems created by the compute-intensive tasks of generating data at the intermediate nodes. Furthermore, an initialization protocol must be designed to extend the efficient prompt generation to data of different modalities. Such protocol would allow the transmitter to distinguish the type of data that is transmitted and learn the appropriate quality function to optimize the communication process. Lastly, GenAI can aid the network layer in facilitating timely and low latency delivery of data by leveraging implicit prompting and can even be used in combination with TPC mechanisms to avoid congestion.

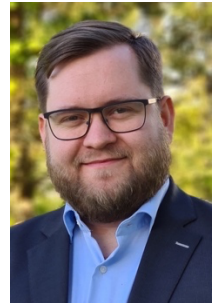

**Mathias D. Thorsager** received the B.Sc. degree in computer engineering and the M.Sc. degree in communication technology from Aalborg University (AAU), Denmark in 2021 and 2023 respectively. He is currently a Ph.D. student at the connectivity section (CNT) of the Department of Electronic Systems, AAU, Denmark. His research interests include beyond 5G and 6G networks, and artificial intelligence and machine learning.

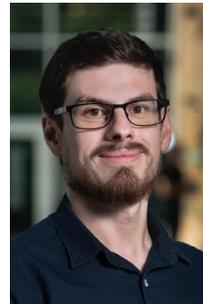

**Israel Leyva-Mayorga** (Member, IEEE) received the B.Sc. degree in telematics engineering and the M.Sc. degree (Hons.) in mobile computing systems from the Instituto Politécnico Nacional (IPN), Mexico, in 2012 and 2014, respectively, and the Ph.D. degree (cum laude and extraordinary prize) in telecommunications from the Universitat Politècnica de València (UPV), Spain, in 2018. He was a Visiting Researcher at the Department of Communications, UPV, in 2014, and at the Deutsche Telekom Chair of Communication Networks, Technische Universität Dresden, Germany, in 2018. He is currently an Assistant Professor at the Connectivity Section (CNT) of the Department of Electronic Systems, Aalborg University (AAU), Denmark, where he served as a Postdoctoral Researcher from January 2019 to July 2021. He is an Associate Editor for IEEE Wireless Communications Letters, a Board Member for one6G, and a representative of AAU in 6G IA. His research interests include beyond-5G and 6G networks, satellite communications, and random and multiple access protocols.

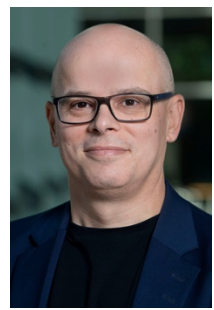

**Petar Popovski** (Fellow, IEEE) received the Dipl.-Ing. and M.Sc. degrees in communication engineering from the University of Sts. Cyril and Methodius, Skopje, and the Ph.D. degree from Aalborg University, in 2005. He is a Professor with Aalborg University, where he heads the Connectivity Section. He is also a Visiting Excellence Chair with the University of Bremen. He has authored the book Wireless Connectivity: An Intuitive and Fundamental Guide. His research interests include wireless communication and communication theory. He received an ERC Consolidator Grant in 2015, the Danish Elite Researcher Award in 2016, the IEEE Fred W. Ellersick Prize in 2016, the IEEE Stephen O. Rice Prize in 2018, the Technical Achievement Award from the IEEE Technical Committee on Smart Grid Communications in 2019, the Danish Telecommunication Prize in 2020, and the Villum Investigator Grant in 2021. He was a Member at Large at the Board of Governors in IEEE Communication Society from 2019 to 2021.

He is the Chair of the IEEE Communication Theory Technical Committee. He was the General Chair of IEEE SmartGridComm 2018 and the IEEE Communication Theory Workshop 2019. He is currently the Editor-in-Chief of IEEE JOURNAL ON SELECTED AREAS IN COMMUNICATIONS.